\documentclass[showpacs,preprintnumbers,amsmath,amssymb,prl,floatfix]{revtex4}
\usepackage{amsmath}
\usepackage{graphicx}
\usepackage[dvipsnames]{xcolor}
\newcommand{\epem}{e^+ e^-}

\begin{document}

\title{The Angular Correlations in the $\epem$ Decay of Excited States in $^8$Be}
\author{A.C. Hayes$^1$, J. Friar$^1$, G.M. Hale$^1$, G. T. Garvey$^{1,2}$}
\affiliation{$^1$Los Alamos National Laboratory, Los Alamos, NM, 87545, USA}
\affiliation{$^2$University of Washington, Seattle, WA, 98195, USA}
\begin{abstract}
Motivated by the recent observation of anomalous electron-positron angular correlations
in the decay  of the 18.15 MeV 1$^+$ excited states in $^8$Be, 
we reexamine in detail the Standard Model expectations for these angular correlations. 
The 18.15 MeV state is above particle threshold, and several multipoles can contribute to its $\epem$ decay.
We present the general theoretical 
expressions for $\epem$ angular distributions for nuclear decay by
C0, C1, C2 M1, E1, and E2 multipoles, and we examine their relative contribution 
to the $\epem$ decay of $^8$Be at 18.15 MeV. We find that this resonance is dominated by M1 and E1 decay,
and  that the ratio of M1 to E1 strength is a strong function of energy.
This  is in contract to the original analysis of the  $\epem$ angular distributions, where the M1/E1 ratio was
 assumed to be a constant over the energy region $E_p=0.8-1.2$ MeV.
We  find that the existence of a `bump' in the measured angular distribution is strongly dependent on the assumed
M1/E1 ratio, with the present analysis finding the measured large-angle contributions to the $\epem$ angular distribution to be
lower than expectation. Thus, in the current analysis we find no evidence for axion decay in the 18.15 MeV resonance region of $^8$Be.
\end{abstract}

\maketitle
\section{Introduction}
Recently, an anomaly has been observed \cite{anomaly} in the electron-positron pair decay of an M1 resonance in $^8$Be. 
In particular, the observed angle between the emitted $e^+$ and $e^-$ pair in the transition of the 18.15 MeV 
1$^+$ resonance to the ground state of $^8$Be deviated significantly from expectation, 
and shows a so-called bump or shoulder at angles greater than 110$^{\circ}$.  
The experiment \cite{anomaly} populated resonances in $^8$Be via the $^7$Li+p reaction and 
observed their decay by detecting $\epem$ pairs. 
Many multipoles can contribute to the observed pair decay with the particular linear combination 
being determined by the nuclear structure of the $^8$Be continuum at each incident proton energy. 
Interference between multipoles can also occur. 
In this experiment, the $\epem$ detectors were set at 90$^{\circ}$ to the incident beam, which tends to minimize the effects of interference. 
For this reason,  we believe the interference effects are small but we could not reliably estimate them without a simulation of the experimental setup and detector responses. 
If the detector resolution and efficiency are symmetric about 90$^\circ$ to the incident beam and sufficient statistics are collected the only effect creating interference between the various multipoles decaying to the ground state would be due to motion of the center of mass of the $\epem$ pair.

\section{Formalism for $\epem$ angular distributions}

 The $\epem$ process is a mechanism in which the nucleus emits a virtual photon which then decays to an $\epem$ pair.
The process is exactly analogous to electron scattering, but with the incident electron of momentum $p$
in the scattering being 
substituted by the outgoing positron in the decay: the momentum
of the positron becomes 
$\bar{p} = - p 
\equiv(\bar{E},\bf{\bar{p}})$. If the 3-momentum of the emitted 
electron is $\bf{p^\prime}$, the momentum transfer is $q=-(\bar{p}+p^\prime)$ 
and $\bf{q}=-(\bf{\bar{p}}+\bf{p^\prime})$. 
Thus, our convention is that ${\bf q}$ is directed into the nucleus. We also ignore Coulomb effects and treat the lepton wave functions as plane waves.

For energies that allow a non-relativistic
treatment of the nuclear physics, the $\epem$ decay by a single nuclear state
is  determined by the  momentum- and nuclear-structure-dependent electromagnetic charge and current operators, whose separate contributions are described throughout this work  by the
structure functions $T^{00}({\bf q})$ and $T^{\perp\perp}({\bf q})$, respectively. 

The first function, $T^{00}(\bf{q})$, is defined in terms of matrix elements of the nuclear charge operator as
\begin{equation}
T^{00}({\bf q}) =\overline{\sum_{f,i}} \mid\langle J_f M_f\mid\rho({\bf q})\mid J_i M_i \rangle  \mid^2,
\label{T00}
\end{equation}
where $\rho(\mathbf{q})=\int d^3x\;\rho(\mathbf{x})\;e^{i\mathbf{q\cdot x}}$ and $\rho(\mathbf{x})$ is the nuclear charge density. The overscore on the summation sign in eq. (\ref{T00})  
means average over initial nuclear spin projections ($M_i)$ and
sum over final ones $(M_f)$.
We assume that  there is no initial-state or target polarization involved.

When expanded in spherical harmonics, the charge operator $T^{00}$ becomes,
\begin{equation}
T^{00}=\frac{4\pi}{2J_{i}+1}\sum_{l\ge 0}\mid\langle J_f\mid\mid C_l\mid\mid J_i\rangle\mid^2,
\label{T001}
\end{equation}
where the normal-parity ($(-1)^l$) charge multipole operator is given by
\begin{equation}
C_{lM}=\int\;d^3x\;\rho({\bf x})j_l(qx)Y_{lM}(\hat{x}).
\label{T002}
\end{equation}
We follow a standard, but not universal, notation that denotes charge multipoles by C$_{\rm J}$, (i.e., C0, C1, C2, etc.). Thus the charge multipoles are distinguished from current multipoles, both electric (E1, E2, etc.) and magnetic (M1, M2, etc.).
In the limit of small $q$, current conservation mandates that the E$_{\rm J}$ are proportional to C$_{\rm J}$ (for $l \ge 1$), which is known as Siegert's Theorem. Only C0 is unique in origin via the charge operator and does not contribute to real photon decay.

The second function, T$^{\perp\perp}(\bf{q})$, is determined by the nuclear current $J^m({\bf q})=\int d^3x\;J^m(\mathbf{x})\;e^{i\mathbf{q\cdot x}}$.
Expanding the exponential in $\ell$ and using the $1^-$ character of the current, leads to the requisite operators, 
\begin{equation}
O^l_{\!\!JM}=\int\;d^3x \,j_l(qx)\left(J_1({\bf x})\otimes Y_l(\hat{x})\right)_{JM}.
\end{equation}
Only $J\ge 1$ contributes to T$^{\perp\perp}$ and we find,
\begin{equation}
T^{\perp\perp}=\frac{4\pi}{2J_i+1} \sum_{J\ge 1} \left(\mid\langle J_f\mid\mid O^{J-1}_J\mid\mid J_i\rangle\mid^2\left(\frac{J+1}{2J+1}\right)+\mid\langle J_f\mid\mid O^J_J\mid\mid J_i\rangle\mid^2\right) .
\label{EM}
\end{equation}
The first set of matrix elements in eq. (\ref{EM}) produces electric (E$_{\rm J}$) transitions and the second 
produces magnetic (M$_{\rm J}$) transitions, with $J\ge 1$ in both cases.

If we denote the $\epem$ decay rate by $\omega_{\epem}$,
we find that the general form of the $\epem$ angular distribution is given by 
\begin{equation}
\frac{d^2\omega_{\epem}}{\!\!\!\!\!\!\!\!dxdy}=\frac{2\alpha^2\bar{p}\,p^\prime\, Q}{\pi q^4}\left\{T^{00}(\mathbf{q})\,
\frac{q^4}{\mathbf{q^4}}\,a^{00}
+T^{\perp\perp}(\mathbf{q})\,a^{\perp\perp}\right\}\;.
\label{big}
\end{equation} 
We denote the angle between the electron and positron by $\theta$, and will use $x$=cos($\theta)$.
Our dimensionless kinematic variable $y$ is similar to that defined  in ref. \cite{anomaly},
$\bar{E}-E^\prime=yQ$, where  $Q$ is the {\bf total} energy of the transition, and
$y$  can lie between -(1-$\frac{2m_e}{Q}$) and 1-$\frac{2m_e}{Q}$; the difference in definitions is discussed in footnote \footnote{In ref. \cite{anomaly}  $\bar{E}-E^\prime=yK$, where $K$ is the sum of the lepton kinetic energies. This difference in definitions has no practical implications.}.
In eq. (\ref{big}) the kinematic functions $a^{00}$ \cite{Friar3} and $a^{\perp\perp}$ are given by
\begin{equation}
a^{00}=\bar{E}E^\prime+\bar{p}p^\prime x -m_e^2,
\end{equation}
and
\begin{equation}
a^{\perp\perp}=\bar{E}E^\prime+ m_e^2-\frac{\mathbf{q\cdot\bar{p}\: q\cdot p^\prime}}{\mathbf{q^2}}=
\bar{E}E^\prime+ m_e^2-\frac{(\bar{p}^2+\bar{p}p^\prime x)\,(p^{\prime 2}+\bar{p}p^\prime x)}{\mathbf{q}^2}.
\label{ans}
\end{equation}
We summarize in Table 1 the definition of the kinematic variables appearing in eq. (\ref{ans}).
\begin{table}[h]
\begin{tabular}{|l| l|}\hline
$\bar{E}=\frac{Q}{2}(1+y)$&$E^\prime=\frac{Q}{2}(1-y)$\\
$\bar{p}=\frac{Q}{2}\left[(1+y)^2-r_Q^2\right]$&$p^\prime=\frac{Q}{2}\left[(1-y)^2-r_Q^2\right]$\\
$q^2=-2(m_e^2+\bar{E}E^\prime- \bar{p}p^{\prime}x)$&$\mathbf{q}^2=\bar{p}^2+p^{\prime 2}+2\bar{p}p^\prime x$\\\hline
\end{tabular}
\caption{The relation between $\bar{E},E^\prime, \bar{p}, p^\prime, q^2, \mathbf{q}^2, Q, x, y$ and $r_Q$, where
$r_Q\equiv\frac{2m_e}{Q}$}.
\end{table}

\subsection{$\epem$ angular distributions for individual multipoles}

The functions 
$T^{00}(\mathbf{q}^2)$ and T$^{\perp\perp}(\mathbf{q}^2)$ in given in terms of reduced nuclear matrix elements. 
We restrict our discussion to four electromagnetic multipoles, namely, 
scalar, vector, pseudoscalar, and tensor,
or equivalently:
C0, E1, M1 and E2. 
We expand $T^{00}(\mathbf{q}^2)$ and T$^{\perp\perp}(\mathbf{q}^2)$ in $ \mathbf{q}^2$ and
keep only leading-order terms.
Using the definition of $T^{00}$ in eqs. (\ref{T001}) and (\ref{T002}), and expanding
the Bessel functions $j_l(qx)$ produces,
\begin{equation}
T^{00}\cong\frac{\mathbf{q}^4S}{36}+\frac{\mathbf{q}^2D}{3}+\frac{\mathbf{q}^4T}{225}\;.
\label{T00big}
\end{equation}
The terms on the right hand size of (\ref{T00big}) determine the C0 ($0^+$), E1 ($1^-$), and E2 ($2^+$) transition rates 
via 3 nuclear-structure constants, $S$, $D$ and $T$, which are  the squared {\bf reduced} monopole, dipole and quadrupole matrix elements, respectively.

The transverse-current structure function, $T^{\perp\perp}$, can be determined in a similar fashion.
Expanding the spherical Bessel functions n eq. (4) results in  magnetic dipole ($M$), electric dipole ($D$) and electric quadrupole ($T$) contributions:
\begin{equation}
T^{\perp\perp}\cong\frac{2\mathbf{q}^2M}{3}+\frac{2Q^2D}{3}+\frac{\mathbf{q}^2Q^2T}{150}.
\label{Tppbig}
\end{equation}
The structure constant $M$ is the square of the reduced magnetic dipole matrix element. The proportionality of the $O^{J-1}_{\!\!JM}$ and the $C_{JM}$ for small $q$ (viz., Siegert's Theorem) was used. Of the four terms that remain (proportional to $S, D, M, T$) we keep the two ($D$ and $M$) that have kinematic coefficients
that are dimensionally equivalent to $(energy)^2$, and ignore the higher-order terms, in common with other discussions.

\subsection{Relation to the photon decay rate}

The $\epem$ transition strength from the resonances of $^8$Be to the ground state can be constrained by the corresponding photon transition strength, as was done by Rose \cite{rose}, who calculated the leading-order $\epem$ decay rate ``per photon.'' We have verified his results. 
In the case of $\gamma$-decay with an outgoing photon of momentum $\mathbf{q}$ in the final state, the decay rate, $\omega_\gamma$, is given by
\begin{equation}
\omega_\gamma=2\alpha Q T^{\perp\perp}(\mathbf{-q})\;.
\label{photon1}
\end{equation}
Thus the squared matrix elements $M$, $D$, and $T$ entering the $\epem$ rates can, in principle, be extracted
from measured $\gamma$-decay rates.
Combining equations (\ref{Tppbig}) and (\ref{photon1}) gives, 
\begin{equation}
\omega_\gamma\cong\frac{4\alpha}{3}\left(Q^3M + Q^3D+\frac{Q^5T}{100}\right)\;.
\label{photon2}
\end{equation}
The squared  E0 matrix element, $S$, must be determined by other means, if needed. We will require only $D$ and $M$. 

\section{Analysis of the photon decay data for $^8$Be from the $^7$Li$(p,\gamma)$ reaction}

The $\gamma$-decays of the $^8$Be resonances of interest have been measured via the analogous $^7$Li$(p,\gamma)$ reaction; the integrated cross section has been measured by  Zahnow {et  al.} \cite{Zahnow} and the $90^{\circ}$ excitation function by Fisher {\it et al.} \cite{Fisher}.
The angular distributions for the emitted photons have been measured by Mainsbridge \cite{Mainsbridge} and 
by Schlueter {\it et  al.} \cite{Schlueter}. 
The resonance energy range of interest is entirely dominated by E1 and M1 photon decay, and 
we used the measured cross sections and shape of the angular distributions in an R-matrix analysis to determine the magnitude of the M1 and E1 contributions to the cross section as a function of energy. 

\begin{figure}[h]
\includegraphics[width=12 cm]{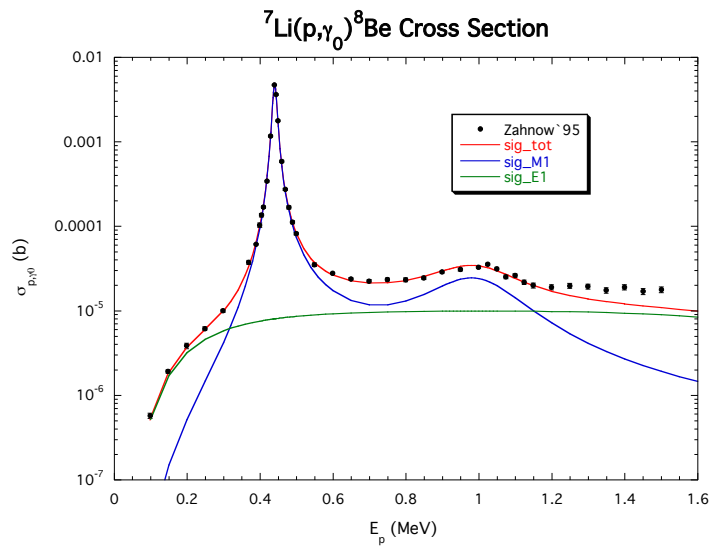}
\caption{The total integrated cross sections for the $^7$Li$(p,\gamma)^8$Be reaction from \cite{Zahnow}. 
The red curve is the result of our R-matrix fit to all of the available cross section and angular distribution data for the reaction.
The blue and green curves show the M1 and E1 contributions to the cross section, respectively.
The small peak centered at $E_p\sim$ 1 MeV is  the 18.15 MeV (1$^+$ T=0) resonance, and the sharp peak at 0.4414 MeV is the 17.64 1$^+$ T=1 resonance in $^8$Be. 
From the R-matrix analysis, as well as from general arguments, the ratio of the magnetic to  electric photon-decay strength varies strongly over the 18.15 MeV resonance. 
\label{Zahnow-data}
}
\end{figure}

\begin{figure}
\includegraphics[width=12 cm]{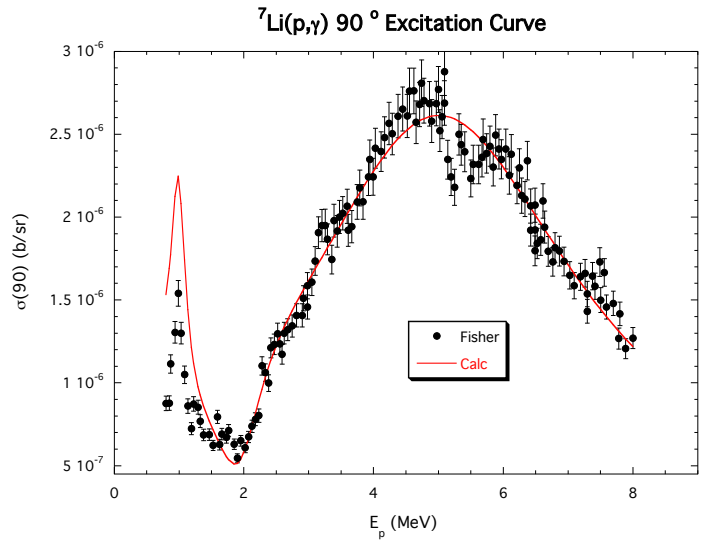}
\caption{ The R-matrix fit to the 90$^{\circ}$ excitation function for the $(p,\gamma)$ reaction. The data are from \cite{Fisher}.
The R-matrix over-estimate of the excitation function may reflect a theoretical  M1 contribution that is too large, but it also reflects the fact that
the Fisher \cite{Fisher} and Zahnow \cite{Zahnow} measurements are not consistent with one another. 
}
\label{fisher-data}
\end{figure}

\begin{figure}[h]
\includegraphics[width=12 cm]{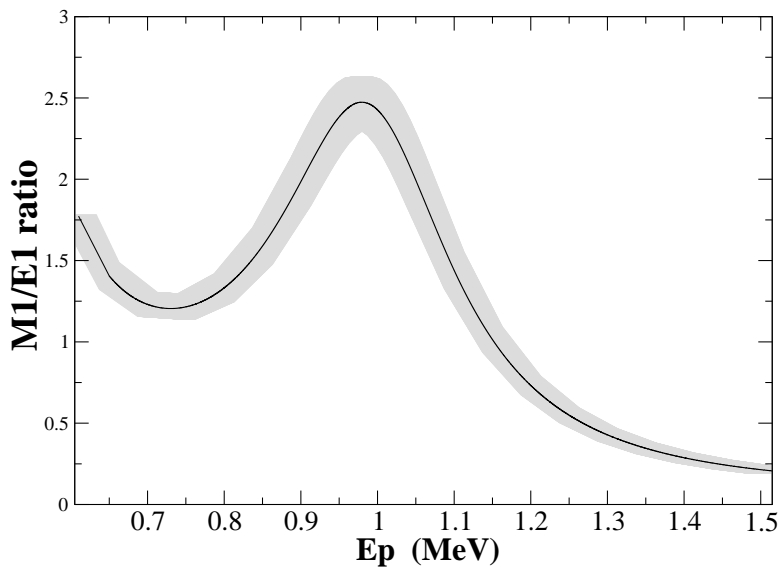}
\caption{ The M1 to E1 photo-absorption cross section ratio from the R-matrix analysis of the  $(p,\gamma)$ cross section and angular distributions.
At the peak of the $E_p$=1.03 resonance, the cross section is approximately determined by an M1+0.45E1 combination, but above $E_p=1.16 $ MeV, the E1 contribution starts to dominate.
}
\label{ratio}
\end{figure}

The R-matrix fit to $(p,\gamma$) cross section data of Zahnow {\it et  al.} is shown in  Fig. \ref{Zahnow-data}.
We find that over the resonance centered at 18.15 MeV ($E_p=0.8-1.2$ MeV) the combination of M1 and E1 multipole strengths 
contributing to the $(p,\gamma)$ reaction, and hence to the $(p,\epem)$ reaction, varies strongly with energy. This 
 is because the M1 strength corresponds to a narrow (138$\pm$6 keV) resonance centered at $E_x=18.15$ ($E_p$=1.03)  MeV, 
whereas the E1 strength comes from the tail of the broad electric dipole structure centered close to $E_p=5$ MeV. This E1 structure is evident in the data
of Fisher {\it et al.}, Fig. \ref{fisher-data}.
The prediction of a significant and broad direct $s$-wave E1 capture to the $^7$Li($p,\gamma$) and $^7$Li($p,\epem$) reactions 
near E$_p$=5 MeV is consistent with the analysis of Barker \cite{barker}.
The ratio of M1 to E1 strength, together with its (shaded) 1-sigma uncertainty,  is shown in Fig. \ref{ratio}. That uncertainty was obtained by scaling the $\chi^2$ from the R-matrix analysis to 1.0 while increasing error bars appropriately. 
Our analysis yields a very different energy dependence to the M1/E1 ratio than that assumed in ref. \cite{anomaly}, where a  constant distribution of 
M1+0.23E1  was assumed over the energy region $E_p=0.8-1.2$ MeV.  However, the large differences in widths of the M1 and E1 resonances
rule out the possibility of the M1/E1 ratio being a constant. 
As discussed below, this strong energy-dependent variation in the M1 and E1 contributions to the $(p,\gamma)$ and $(p,\epem)$ reactions has significant implications 
for any axion search close to the $E_p$=1.03 MeV resonance in $^8$Be.

The R-matrix analysis over-estimates the 90$^{\rm o}$ excitation function in the region of the 18.15 MeV resonance, which may reflect 
an over-estimate of the M1 strength in this region.
However, the problem arises because the Zahnow and Fisher data sets are not consistent with one another at this energy. 
This can be understood by the following considerations:  
If the angular distributions are described by a  
Legendre polynomial expansion up through order $L=2$, then the ratio of the 90$^{\circ}$  to the integrated cross section is determined by, 
\begin{equation}
R=\frac{4\pi \sigma(90^{\circ})}{\sigma_{\rm int}}=1-\frac{1}{2}\frac{a_2}{a_0},
\end{equation} 
in terms of the Legendre coefficients $a_L$.  The maximum value of $\frac{a_2}{a_0} $ is obtained when only the transition $^3P_1(p+^7$Li)$\rightarrow {\rm M1}(\gamma+^8$Be) is allowed, in which case the ratio has the value $\frac{1}{2}$.  The presence of any other transition in the capture reaction, and in particular an E1 transition, dilutes this ratio so that $\frac{a_2}{a_0}<\frac{1}{2}$, meaning that the minimum value of $R$ is 0.75.  The value of $R$ obtained from the R-matrix fit is $0.82 \pm 0.029$, reflecting the non-negligible amount of E1 cross section that contributes near the 1-MeV resonance (see Fig. \ref{Zahnow-data}).  However, using the experimental values of the cross sections near 1 MeV from Fisher and Zahnow gives $R_{\rm exp}=0.59 \pm 0.037$, which is well below the minimum possible value of $R$, and inconsistent with the calculated value within the uncertainties.  Therefore, either the 90$^{\circ}$ cross section of Fisher is low, or the integrated cross section of Zahnow is high, at these energies.  Given that Zahnow's measurement covers the range of both $1^+$ (mostly) M1 resonances continuously, and it was done with five times better energy resolution than that of Fisher's, we tend to favor the explanation that the Fisher measurement is low in the peak of the 1-MeV resonance (see Fig. \ref{fisher-data}). 


\section{The electron-positron angular distributions}
\begin{figure}[h]
\includegraphics[width= 8 cm]{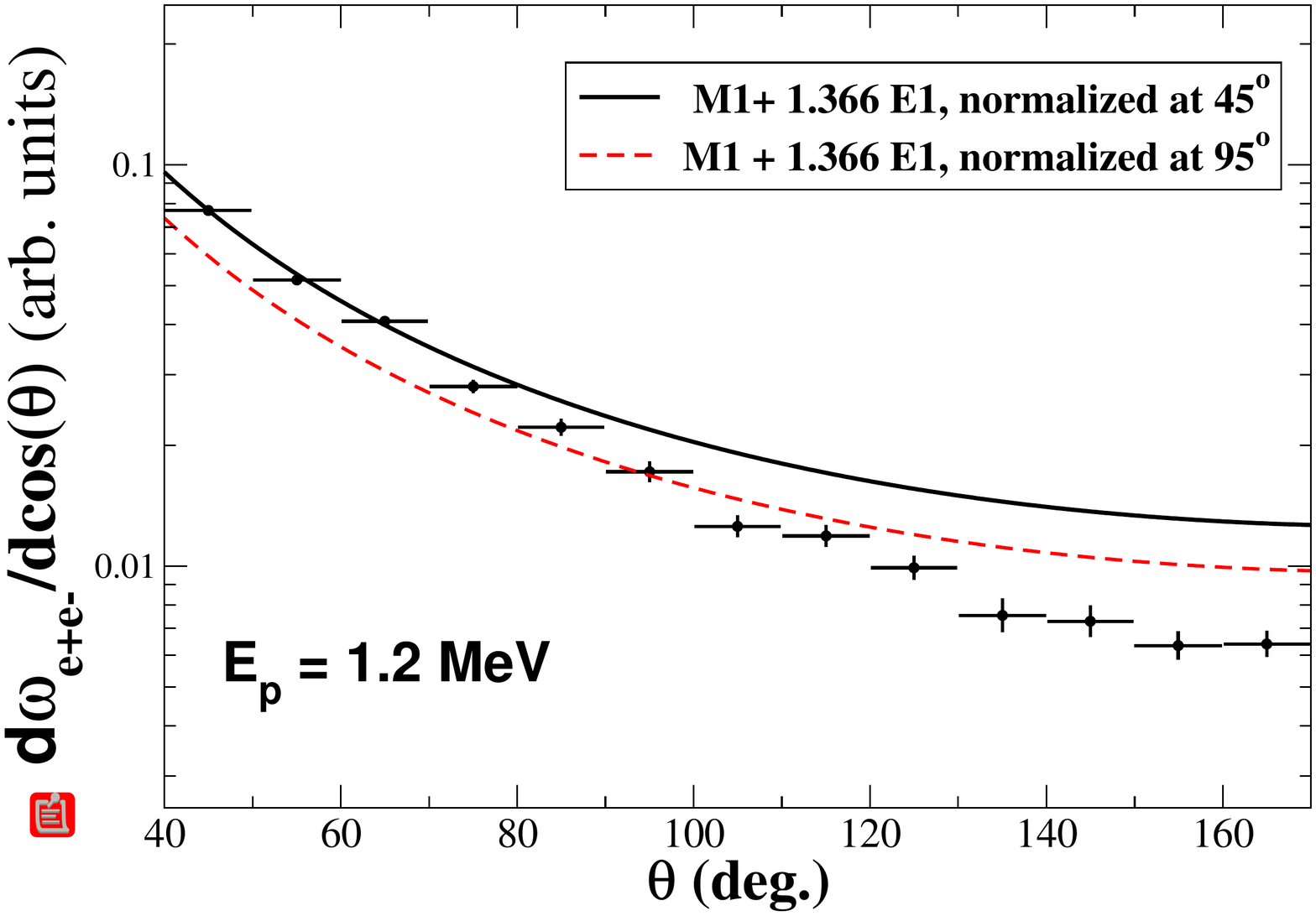}
\includegraphics[width= 8 cm]{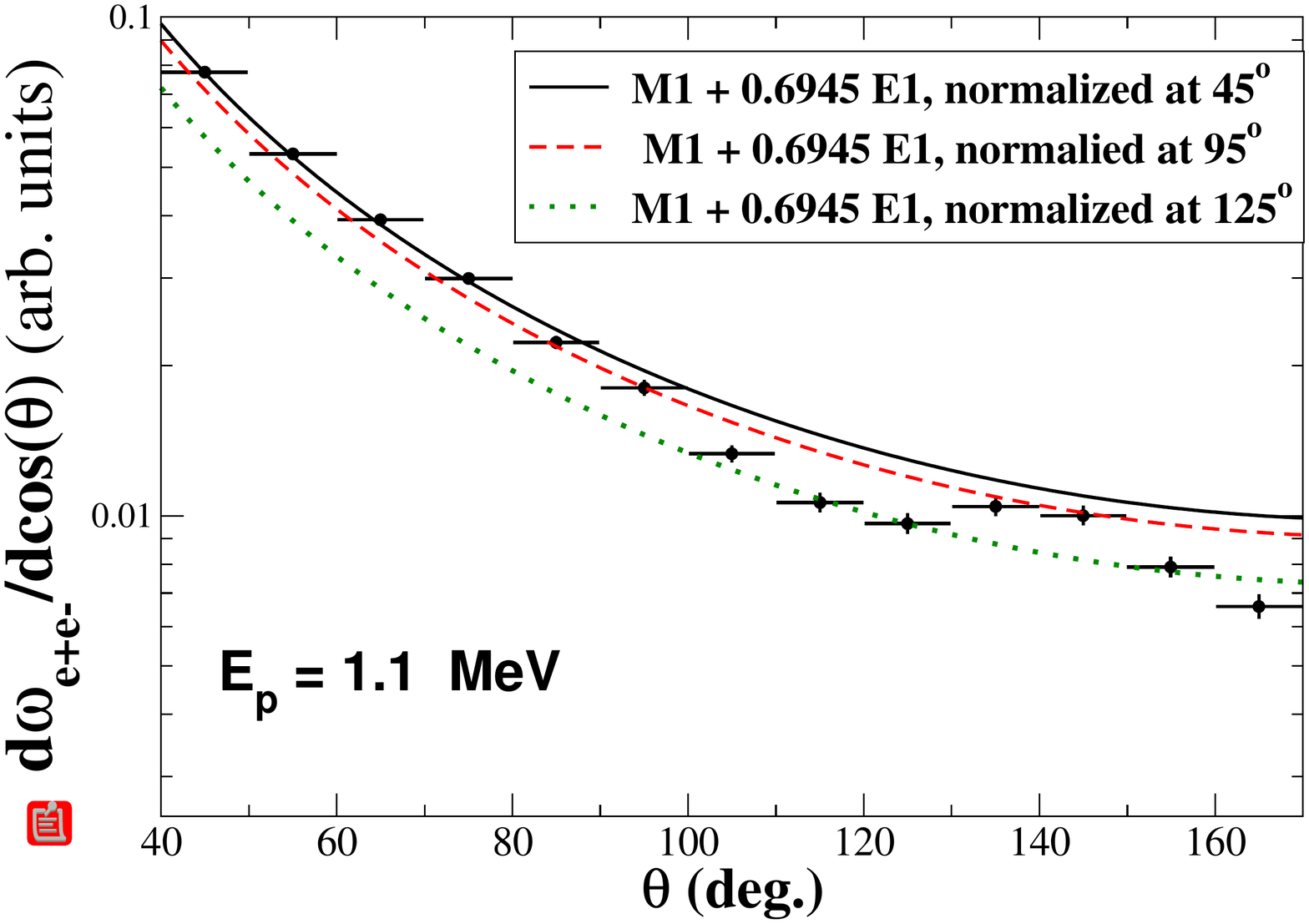}
\includegraphics[width= 8 cm]{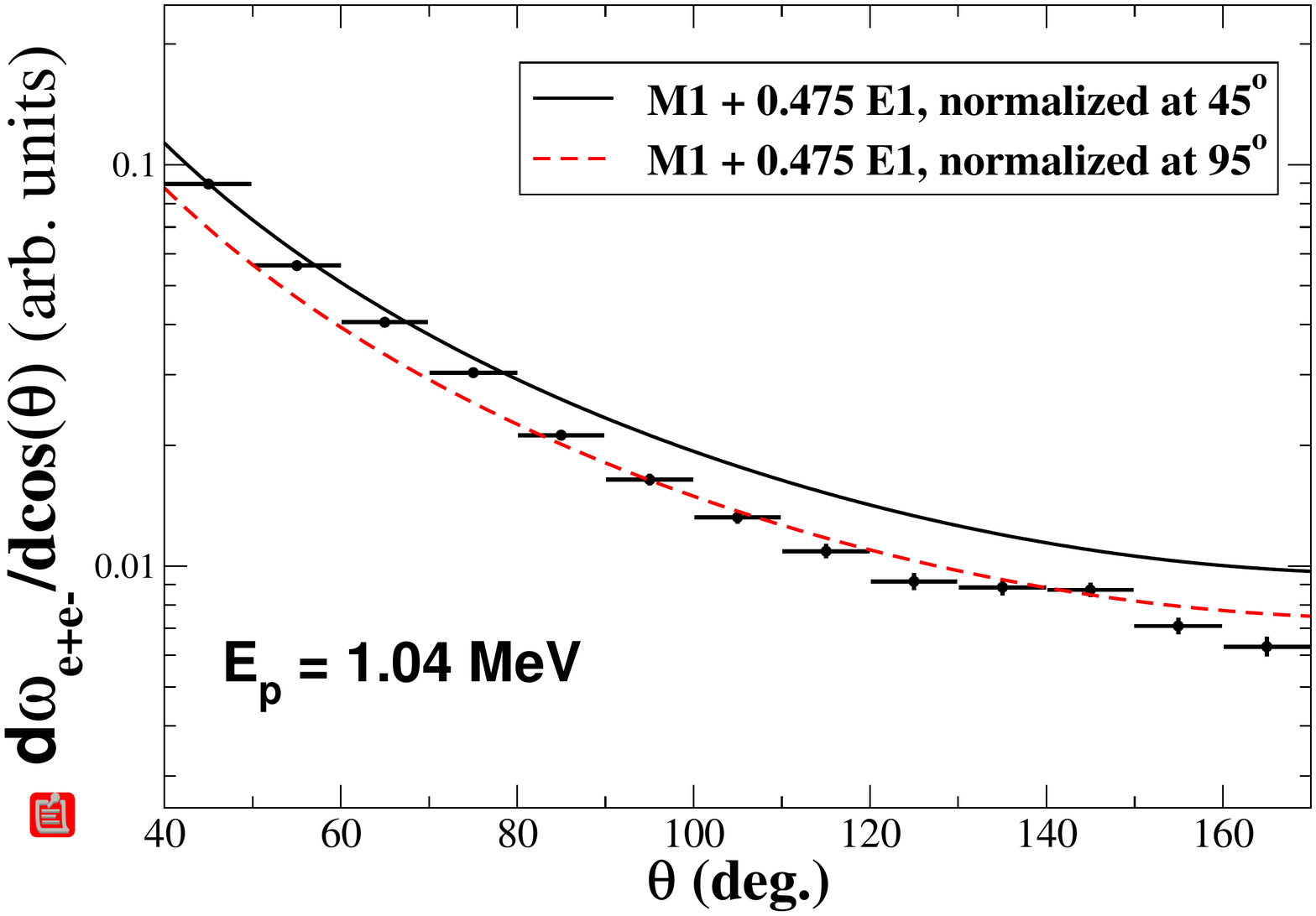}
\includegraphics[width= 8 cm]{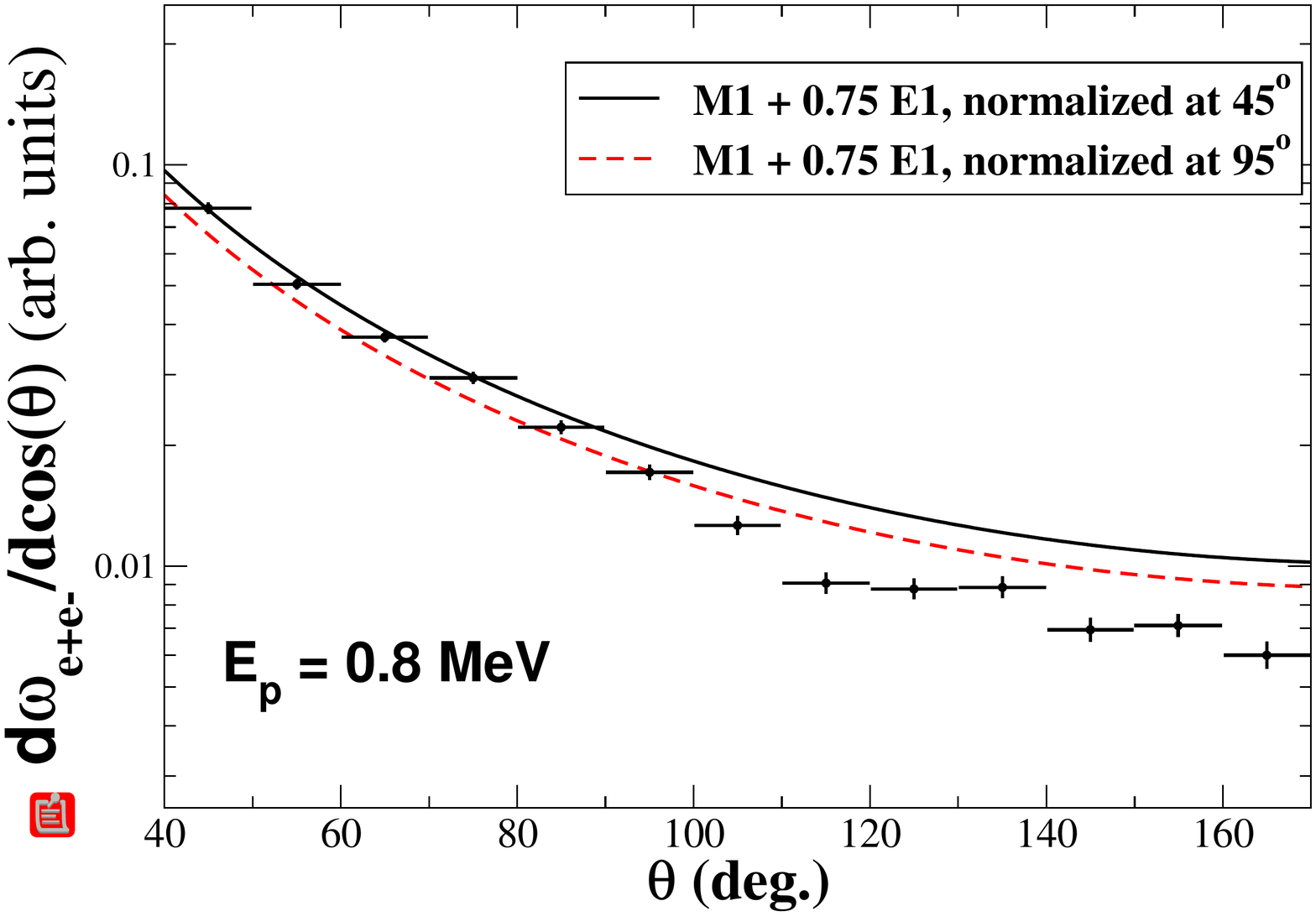}
\caption{The angular distributions for four proton beam energies that sweep across the 18.15 MeV resonance of $^8$Be.
At each energy we take the relative M1 and E1 contribution to the angular distributions from our R-matrix analysis, as summarized in Fig. \ref{ratio}.
If we normalize theory and experiment at forward angles, where the angular distribution is largest, we  find the 
measured angular distributions are too low at backward angles.
The theoretical angular distributions do not show any evidence of a `bump' and the bump seen in experiment lies below the expected 
angular distribution. 
In this sense, the experiment does not appear consistent with  evidence for $\epem$ decay of a new particle. 
For the sake of displaying the bump more clearly, we also show the case where theory is normalized to experiment at 125$^{\circ}$  for
$E_p$=1.1 MeV, but this causes the comparison between theory and experiment at forward angles to be problematic. }
\label{y2}
\end{figure}

The $\epem$ angular distribution can now be calculated by numerically integrating eq. (\ref{big}) 
with respect to the variable $y$. To be consistent with  
ref. \cite{anomaly}, we take the limits of integration to be $0.5>y'>-0.5$, where
$y' =(\bar{E}-E^\prime)/(\bar{k}+k^\prime)$, and $\bar{k}$ and $k^\prime$ are the kinetic energies of the positron and electron. 
We find reasonably good agreement with the measured \cite{anomaly} angular distribution for the 6.05 MeV state in $^{16}$O and the 1$^+$ resonance in $^8$Be excited by protons of energy $E_p=0.441$ MeV, but not for the 1$^+$ resonance excited by proton energies between $E_p=0.8-1.2$ MeV. 
The results for the latter resonance are shown in Fig. \ref{y2}, where they are compared with the measured angular distributions of \cite{anomaly}.
At each proton energy the M1/E1 ratio is taken from our R-matrix analysis. 
We find that the measured angular distributions fall off faster that the theoretical predictions at large angles for proton energies between $E_p=0.8-1.2$.
This effect reflects the fact that the pure M1 angular distribution falls off faster at large angles that does the pure E1 angular distribution, 
and that the current analysis predicts considerably more E1 strength contributing to the resonance that that assumed in \cite{anomaly}.
The lowest E1/M1 ratio in the proton energy range $0.8-1.2$ MeV region is found to be 0.4.
As a result, the bumps observed in the experimental angular distributions fall
below the Standard Model nuclear physics predictions. In this sense, the current analysis does not support 
the measured angular distribution being interpreted as evidence for the decay of a new axion particle.

\begin{figure}[h]
\includegraphics[width= 8 cm]{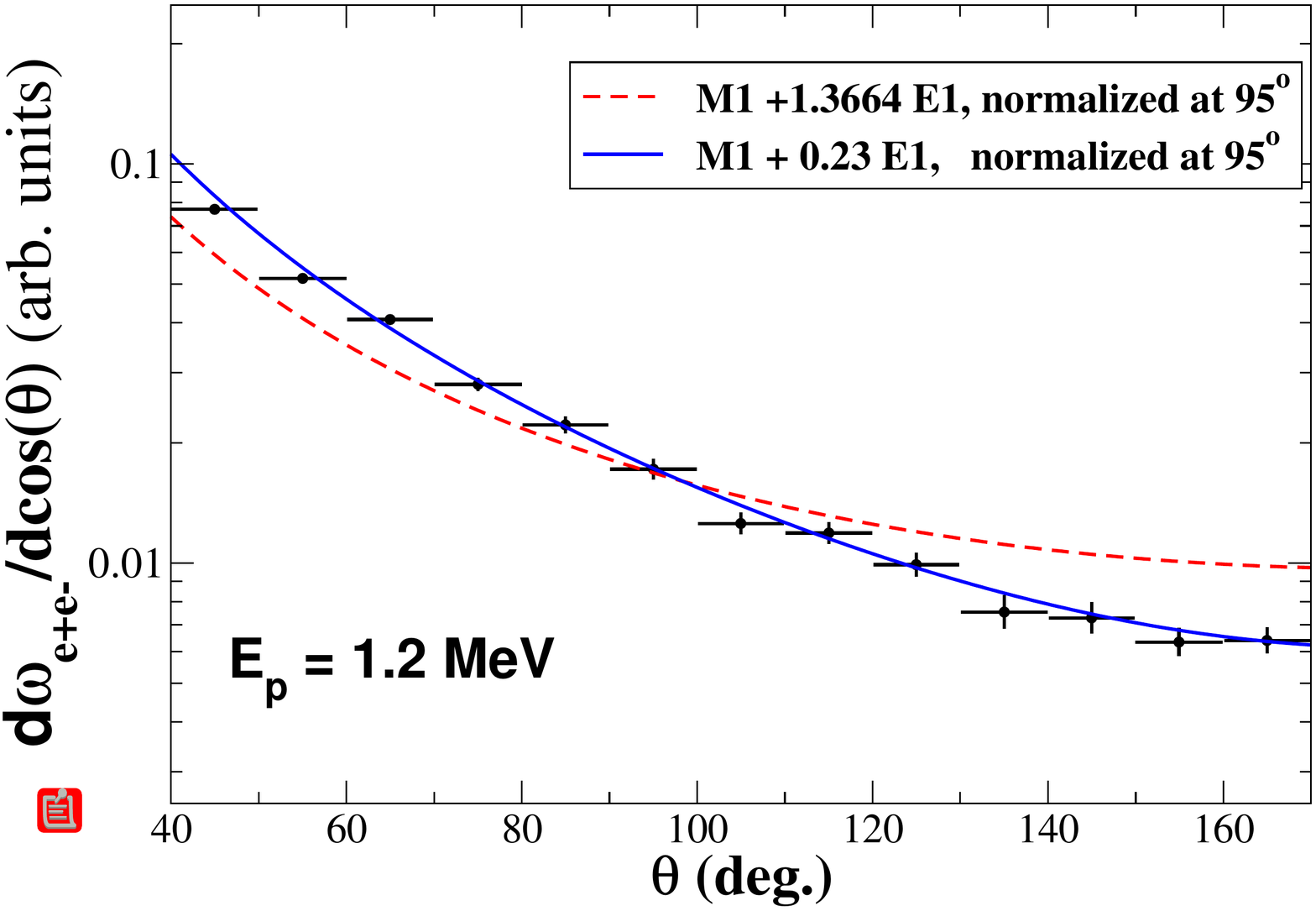}
\includegraphics[width= 8 cm]{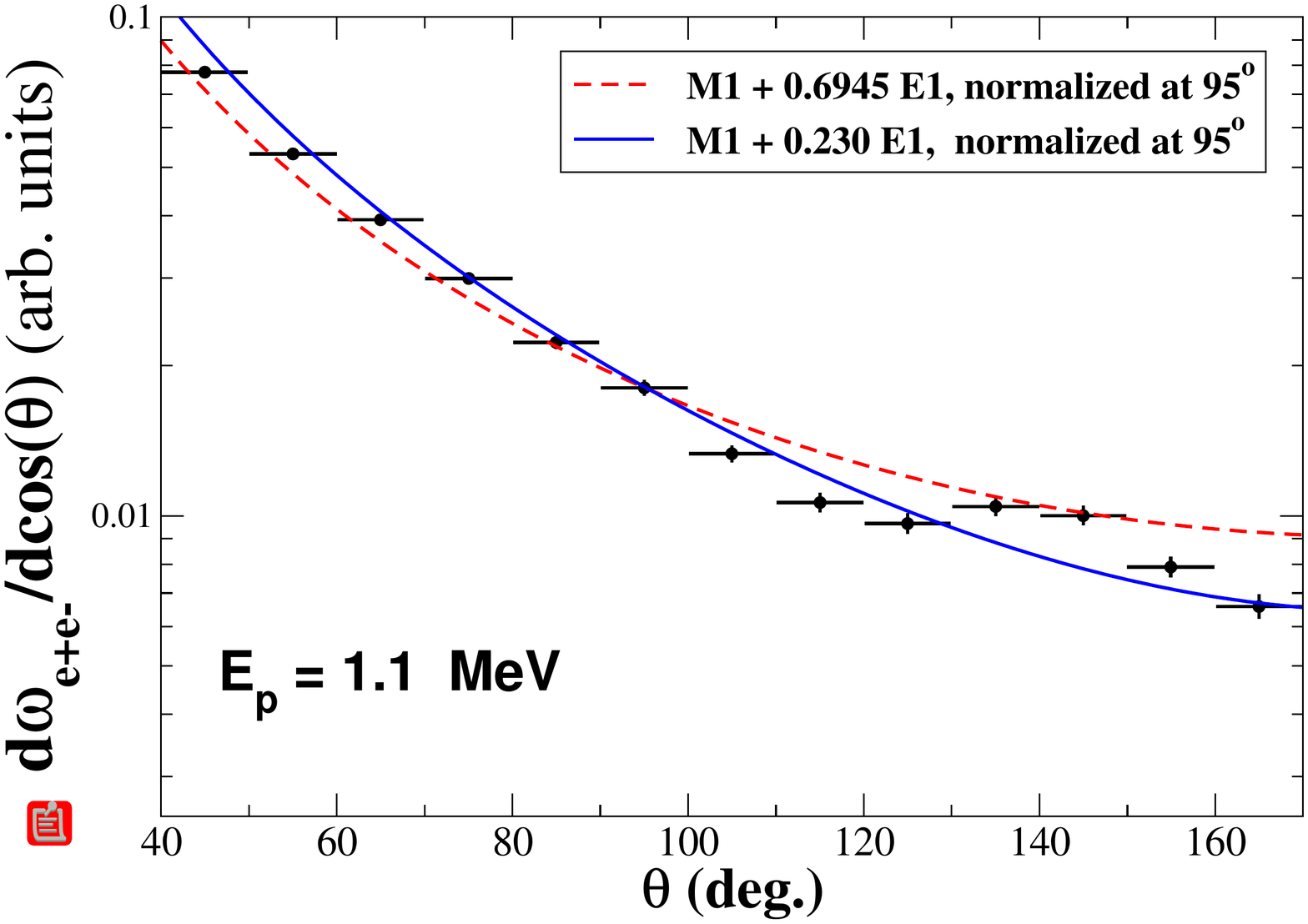}
\includegraphics[width= 8 cm]{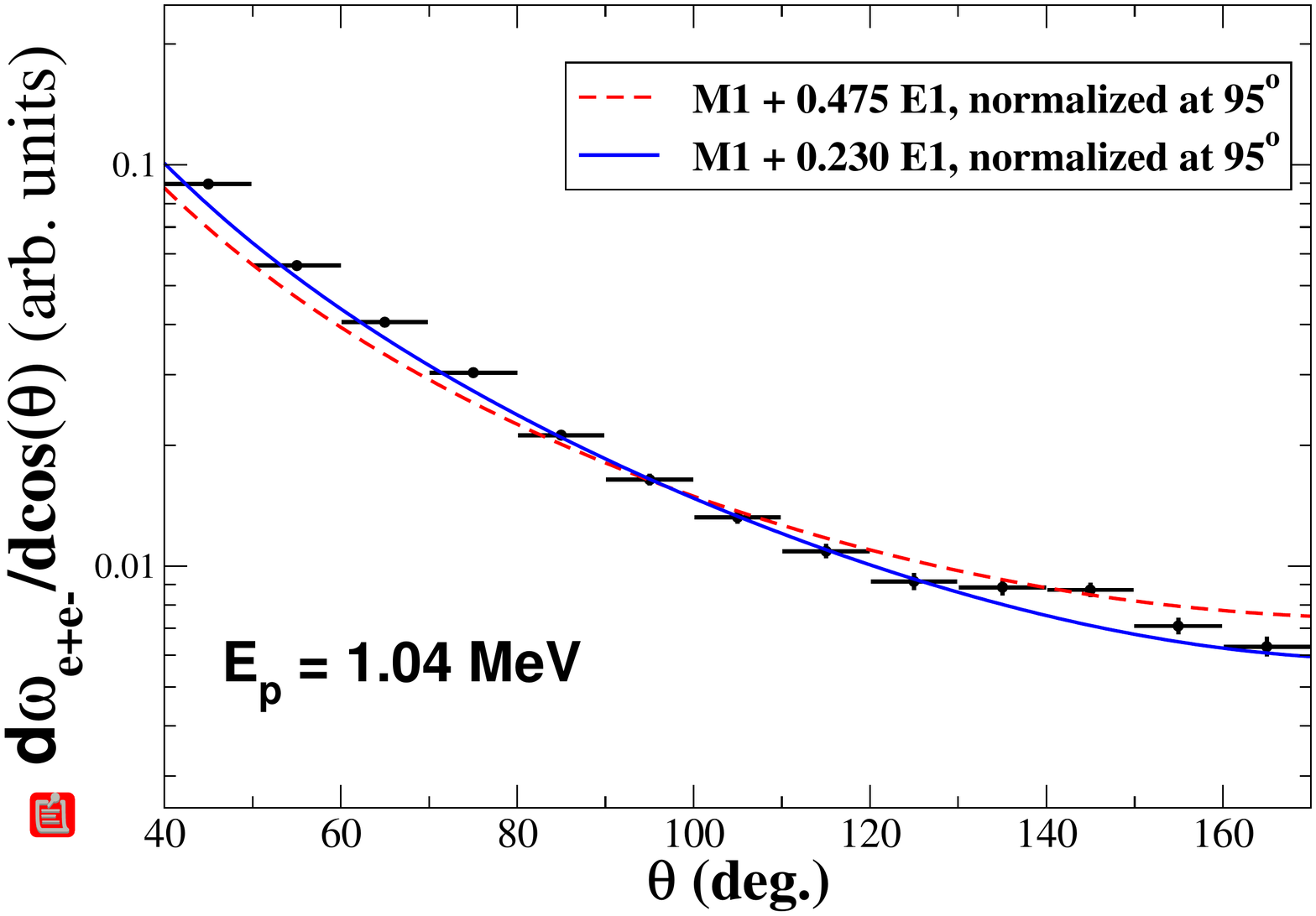}
\includegraphics[width= 8 cm]{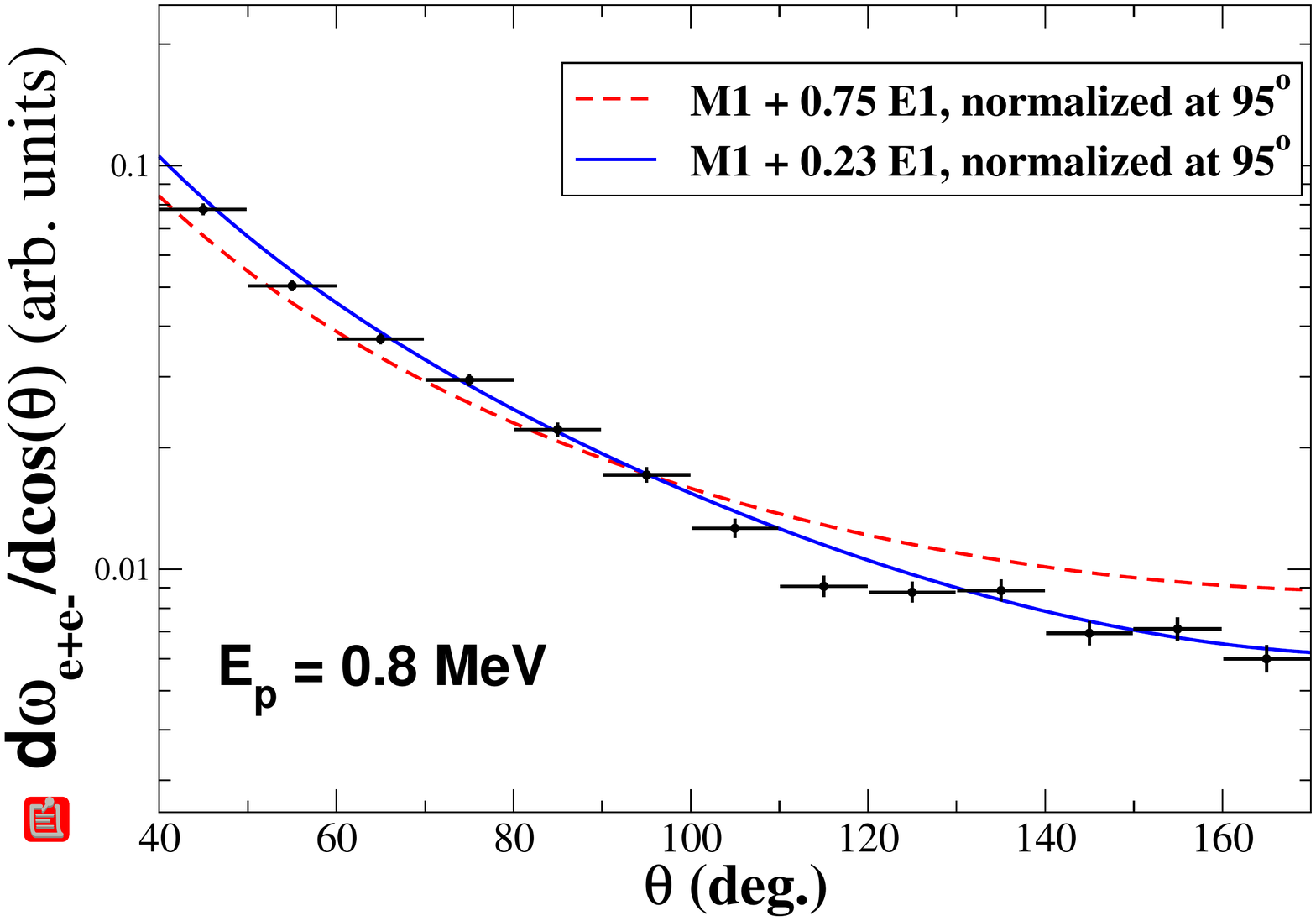}
\caption{The the same as Fig. \ref{y2}, but with a comparison to the angular distribution
 assumed in ref. \cite{anomaly}, i.e. M1 + 0.23E1 (the blue curve). The experimental data do show a `bump' relative to the blue curve, but not relative to the expectations based on the $(p,\gamma$) reaction (the dashed red curve).}
\label{y3}
\end{figure}

Finally, in Fig. \ref{y3} we compare the measured angular distributions with the M1/E1 ratio assumed in ref. \cite{anomaly}, i.e. a constant ratio of M1+0.23E1. Under this assumption, the angular distributions at $E_p=1.04$ MeV and 1.1 MeV do show a bump above the expected angular distributions. 
Thus, we conclude that the evidence of a new axion particle being emitted from the 18.15 MeV resonance in $^8$Be seems to be strongly dependent on the assumptions made about the nuclear structure of this resonance. 

\section{conclusion}
        We have derived expressions for angular correlations for nuclear decay by $\epem$ decay, 
that are completely consistent with earlier expressions derived by Rose \cite{rose}. 
To establish the appropriate combination of multipoles that apply in the region of the 18.15, 1$^+$ MeV resonance in $^8$Be 
we carried out an R-matrix analysis of the available cross sections and angular distributions for the $^7$Li$(p, \gamma)$ reaction.
 We find that the resonance region is entirely dominated by the M1 and E1 multipoles, but that the M1 to E1 ratio varies significantly with energy, 
being a maximum of M1+0.455 E1 at the peak of the resonance (E$_p$=1.03 MeV) and 
dropping to where E1 dominates over M1 above Ep = 1.16 MeV. 
This is in strong contrast to the assumptions of ref. \cite{anomaly}, 
that assumed a constant value of M1+0.23 E1 over the range E$_p$  = 0.8 -1.2 MeV.

The existence of a `bump' in the $\epem$ angular correlations at large opening angle is found to be strongly 
dependent on the assumptions made about the M1/E1 ratio in the 
energy interval associated with the E$_p$=1.03 MeV 1$^+$ resonance. 
The current analysis indicates that the measured angular correlations fall off too rapidly with angle, 
falling below the R-matrix expectations at angles greater than 100$^\circ$. 
In contrast, the nuclear structure assumptions made in the analysis applied in ref. \cite{anomaly} 
would require the measured angular distributions at resonance to fall even more rapidly, 
creating the surplus events at large angles 
thus providing evidence for a new axion-like particle. 
At the least, a detailed re-measurement of the energy dependence of the M1 and E1 yields in the $(p,\gamma)$ reaction 
in this energy range would be needed before the unexpected observation of non-Standard Model particle should be claimed.

\end{document}